\documentclass[a4paper,10pt,aps,pre,twocolumn,amsmath,amssymb]{revtex4}
\usepackage{graphicx,color}

\newcommand{\ahum}[1]{``#1''}
\newcommand{\eq}[1]{Eq.~(\ref{#1})}
\newcommand{\fig}[1]{Fig.~\ref{#1}}

\newcommand{\olcite}[1]{Ref.~\cite{#1}}
\newcommand{\avg}[1]{\langle #1 \rangle}

\newcommand{\tc}{T_{\rm c}}
\newcommand{\etal}{{\it et al.}}

\begin{document}

\title{Universality class of a displacive structural phase transition in two 
dimensions}

\author{Richard L. C. Vink}

\affiliation{Institute of Materials Physics, University of Goettingen, 
Friedrich-Hund-Platz~1, D-37077 Goettingen, Germany}

\date{\today}

\begin{abstract} The displacive structural phase transition in a two-dimensional 
model solid due to Benassi \etal~[Phys.~Rev.~Lett.~{\bf 106}, 256102 (2011)] is 
analyzed using Monte Carlo simulations and finite-size scaling. The model is 
shown to be a member of the two-dimensional six-state clock model universality 
class. Consequently, the model features two phase transitions, implying the 
existence of three thermodynamically distinct phases, namely, a low-temperature 
phase with long-ranged order, an intermediate critical phase with power-law 
decay of correlations, and a high-temperature phase with short-ranged order. \\

\color{red}{{\bf Please note: This is a major revision compared to version 1 (24 
Aug 2018) 
of this manuscript! The reader should use the present version, or directly 
consult the journal reference: 10.1103/PhysRevE.98.062109}} \end{abstract}

\maketitle

\section{Introduction}

Complex oxide solids are known to exhibit structural phase 
transitions~\cite{citeulike:5298220}. These transitions are relevant for 
applications, as material properties below and above the transition typically 
differ, but are also interesting in their own right, regarding the fundamentals 
of solid phase behavior. Consequently, these transitions have received 
considerable attention, including increasingly, by means of computer simulation. 
An example is the study of \olcite{citeulike:13839098}, where a simple particle 
model is proposed featuring a displacive structural phase transition. We shall, 
in what follows, refer to this model as the BVST model (after the first letter 
of each author's surname) and focus exclusively on its {\it equilibrium} phase 
behavior.

Despite the apparent simplicity of the BVST model, its equilibrium phase 
behavior is far from trivial. Indeed, the authors already announce the 
possibility of a Kosterlitz-Thouless (KT) transition, citing 
\olcite{joseRenormalizationVorticesSymmetrybreaking1977}. Further inspection of 
\olcite{joseRenormalizationVorticesSymmetrybreaking1977}, as well as following 
up on valuable comments made by an anonymous referee of an earlier draft of this 
manuscript~\cite{vinkCriticalBehaviorDisplacive2018a}, suggest that the BVST 
model could be in the universality class of the two-dimensional $q=6$ clock 
model. If this is the case, the model should, in fact, undergo two separate 
phase transitions, and, consequently, support three thermodynamically distinct 
phases~\cite{joseRenormalizationVorticesSymmetrybreaking1977}. The purpose of 
the present paper is to verify, via Monte Carlo (MC) simulations and finite-size 
scaling, whether this scenario applies.

As it turns out, our simulations strikingly show that the BVST model is a member 
of the $q=6$ clock model universality class. The existence of two phase 
transitions, implying three phases, is clearly visible. In addition, critical 
exponents obtained using finite-size scaling, are consistent with those of the 
$q=6$ clock model. In what follows, we will present the analysis leading to 
these results. The outline is as follows: We first describe the BVST 
model~\cite{citeulike:13839098}, followed by a brief summary of the $q=6$ state 
clock model. We then present our MC data, followed, in Section~IV, by a 
discussion and summary. The details of the MC methods used in this work are 
provided in the Appendix.

\section{Models}

\subsection{BVST model}

The BVST model~\cite{citeulike:13839098} provides a simple description of a 
material exhibiting a structural phase transition. It qualitatively resembles a 
system whereby, during the transition, one of the particle species remains fixed 
(as do, e.g.~Ba atoms in the case of $\rm BaTiO_3$~\cite{citeulike:5298220}). 
The fixed species is assumed to provide an underlying lattice structure, as well 
as to give rise to a (static) external field acting on the mobile species. It is 
assumed there is only one type of mobile particle species, and the total number 
of these particles is denoted~$N$. In addition, there is a pair interaction 
between the mobile species, described in the form of permanent anharmonic bonds. 
The total energy of the system is thus given by $E = \sum_{[ij]} u_{ij} + 
\sum_{i=1}^N h_i$, where $[ij]$ is a sum over bonded pairs of particles $i$ and 
$j$, $u_{ij}$ the corresponding bond energy, and $h_i$ the external field acting 
on particle~$i$.

The underlying crystal structure is assumed to be a hexagonal lattice, i.e.~the 
model is purely two-dimensional. The nearest neighbor distance between lattice 
points is denoted~$a$. To each lattice position $\vec{R}_i$, a single particle 
is assigned $(i=1,\ldots,N)$. The {\it displacement} of particle $i$ from 
its lattice position $\vec{R}_i$ is denoted $\vec{r}_i=(x_i,y_i)$. During the 
simulations, the particle displacements $\vec{r}_i$ are allowed to fluctuate, 
but the lattice positions $\vec{R}_i$ remain fixed.

Each particle is bonded to its six nearest neighbors by means of an anharmonic 
spring. The spring energy is given by $u_{ij} = b_2 (d_{ij}-a)^2 + b_4 
(d_{ij}-a)^4$ with $d_{ij} = |\vec{R}_j + \vec{r}_j - \vec{R}_i - \vec{r}_i|$ 
the distance between the two particles. The bonds ($3N$ in total) are assigned 
once at the start of the simulation. During the simulations, there is no 
breaking of bonds, nor the formation of new bonds.

The external field acting on particle $i$ is defined in terms of its 
displacement $\vec{r}_i$ as follows:
\begin{equation*}
 \frac{h_i}{\epsilon} = 1 + \left( \frac{r_i}{a_0} \right)^4 - 
 2 \left[ 
  (3 \cos\theta_i - 4\cos^3 \theta_i) \frac{r_i}{a_0}  
 \right]^2 \quad,
\end{equation*}
with $r_i=\sqrt{x_i^2+y_i^2}$ and $\cos\theta_i=x_i/r_i$. Due to the external 
field $h_i$, each lattice position $\vec{R}_i$ is \ahum{surrounded} by six local 
energy minima, at coordinates $\vec{R}_i + a_0 (\cos \pi \lambda/3, \sin \pi 
\lambda/3)$, with $\lambda=1,2,3,4,5,6$. A single particle can thus minimize its 
field energy by selecting one of the surrounding minima. In order to 
simultaneously minimize the bond energy additionally requires that all particles 
choose the same $\lambda$, which leads to the ground state of the model, where 
the total energy $E=0$. Hence, upon lowering the temperature, one expects 
ordering to occur, whereby the particles collectively choose the same value of 
$\lambda$, reminiscent of a displacive structural phase transition. As will be 
shown later, the transition to the ordered (low-temperature) state proceeds via 
two phase transitions.

In what follows, $b_2 = 28.32U/a^2$, $b_4=784.35U/a^4$, $\epsilon=0.2U$, and 
$a_0/a=0.05$, which are the values of the original 
reference~\cite{citeulike:13839098}. The lattice constant $a$ will be our unit 
of length, and temperature will be expressed in units of $U/k_B$, with $k_B$ the 
Boltzmann constant. We use rectangular $L_x \times L_y$ simulation cells with 
periodic boundary conditions in both dimensions. To prepare the initial 
hexagonal lattice, we take as unit cell a $l_x \times l_y$ rectangle, with 
$l_x=a$ and $l_y=\sqrt{3}a$. The unit cell contains two lattice sites, at 
coordinates $(0,0)$ and $(l_x/2,l_y/2)$. This unit cell is then replicated $2n$ 
times in the $x$-direction, and $n$ times in the $y$-direction, with integer 
$n$. Consequently, $N = 4n^2$, $L_x=2an$, $L_y=\sqrt{3} an$, and it is ensured 
that, irrespective of $n$, all our simulation cells have the same aspect ratio 
$L_x/L_y$.

\subsection{$\bf q=6$ clock model}

The (two-dimensional) $q$-state clock model considers a two-dimensional space 
lattice (e.g.~square, hexagonal) of sites $i=1,\ldots,N$. To each lattice 
site~$i$, a discretized direction is attached, expressed via the angle $\theta_i 
= 2\pi n_i/q$, with integer $n_i=1,\ldots,q$. The energy $E = -J \sum_{[ij]} 
\cos(\theta_i-\theta_j)$, coupling parameter $J>0$, and sum over pairs of 
nearest neighbors. The case $q=2$ is the Ising model; $q \to \infty$ the 
XY-model. For $q=6$, the clock model features two phase 
transitions~\cite{joseRenormalizationVorticesSymmetrybreaking1977}, at 
temperatures $T_1$ and $T_2$, respectively (we set $T_1<T_2$ in what follows). 
The model thus supports three, thermodynamically distinct, phases. The phases are 
characterized by the decay of the angular correlation function, $G(r) = \avg{ 
\sum_{r_{ij}=r} \cos(\theta_i - \theta_j)} / N_r$, where the sum is over all 
pairs of sites $i-j$ separated by a distance $r_{ij}=r$, $N_r$ the number of 
such pairs, and $\avg{\cdot}$ a thermal average.

In the high-temperature phase, $T>T_2$, the correlations decay exponentially to 
zero, $G(r) \propto e^{-r/\xi}$, $\xi$ being the correlation length. The phase 
is disordered: There is no alignment of the angular directions over distances 
beyond $\sim \xi$, implying that the order parameter $\Delta=0$. Provided the 
simulation box dimensions $L_x,L_y > \xi$, one expects only negligible 
finite-size effects in simulation data.

In the low-temperature phase, $T<T_1$, there is long-ranged order, with a 
macroscopic fraction of the site orientations \ahum{pointing} in the same 
direction (which can be any one of the $q$ possibilities). Consequently, the 
order parameter $\Delta>0$. The correlation function still decays exponentially, 
but to a finite value, $G(r)-G_\infty \propto e^{-r/\xi}$, with $G_\infty>0$. 
Provided $L_x,L_y > \xi$, finite-size effects in simulation data are again 
negligible.

The intermediate phase, $T_1<T<T_2$, is a {\it critical} phase, where the 
correlations decay as a power law, $G(r) \propto r^{-\eta}$, implying that $\xi$ 
is infinite. Power law decay of correlations also implies that, in the 
thermodynamic limit, the susceptibility $\chi$ is infinite, and the order 
parameter $\Delta=0$. Since $L_x,L_y \ll \xi$ is now unavoidable, finite-size 
effects in simulation data are strong!

We emphasize that the correlation length $\xi$, the plateau value $G_\infty$, 
and the exponent $\eta$ are functions of temperature. According to theoretical 
predictions, $\eta(T_2) \equiv \eta_2 =1/4$, $\eta(T_1) \equiv \eta_1 
=4/q^2=1/9$~\cite{joseRenormalizationVorticesSymmetrybreaking1977}. Simulation 
estimates~\cite{challaCriticalBehaviorSixstate1986, 
tomitaProbabilitychangingClusterAlgorithm2002} are close to these values, though 
not in perfect agreement.

The consensus is (but do note discussions in 
Refs.~\cite{lapilliUniversalityAwayCritical2006, hwangSixstateClockModel2009, 
baekCommentSixstateClock2010}) that both transitions are of the 
KT-type~\cite{kosterlitz.thouless:1973, kosterlitz:1974}, implying exponential 
growth of $\xi$ upon approach of the critical phase:
\begin{equation}
\label{eq:kt}
\xi(T)  \propto \begin{cases}
 e^{a_1 t_1^{-1/2}}, \quad t_1 \equiv \frac{T_1-T}{T_1} & (0 < t_1 \ll 1), \\
 \infty & (T_1 \leq T \leq T_2), \\
 e^{a_2 t_2^{-1/2}}, \quad t_2 \equiv \frac{T-T_2}{T_2} & (0 < t_2 \ll 1). \\
\end{cases}
\end{equation}
For the XY-model at finite temperature, only the transition at $T_2$ remains, 
for which $a_{XY} \sim 1.5$~\cite{kosterlitz:1974}. Computer 
simulations~\cite{challaCriticalBehaviorSixstate1986} show that this value is 
compatible with the $q=6$ clock model as well, for {\it both} transitions: 
$a_1=a_2 \sim 1.54$.

The specific heat $c_V$ of the $q=6$ clock model always remains finite. However, 
the variation of $c_V$ with temperature does reveal two maxima, one occurring 
close to $T_1$, the other closer to $T_2$. This property is convenient in 
simulations to provide first evidence of two phase 
transitions~\cite{challaCriticalBehaviorSixstate1986}.

\section{Results}

In what follows, moderate system sizes $L \equiv L_x = 30-80$ are used to study 
the BVST model, see Appendix for details about the MC methods. This approach is 
in line with \olcite{orkoulas.panagiotopoulos.ea:2000}, where it was noted that 
moderate system sizes, in combination with finite-size scaling methods, can 
yield a very adequate description of the phase behavior in the thermodynamic 
limit.

\subsection{Specific heat}

\begin{figure}
\begin{center}
\includegraphics[width=\columnwidth]{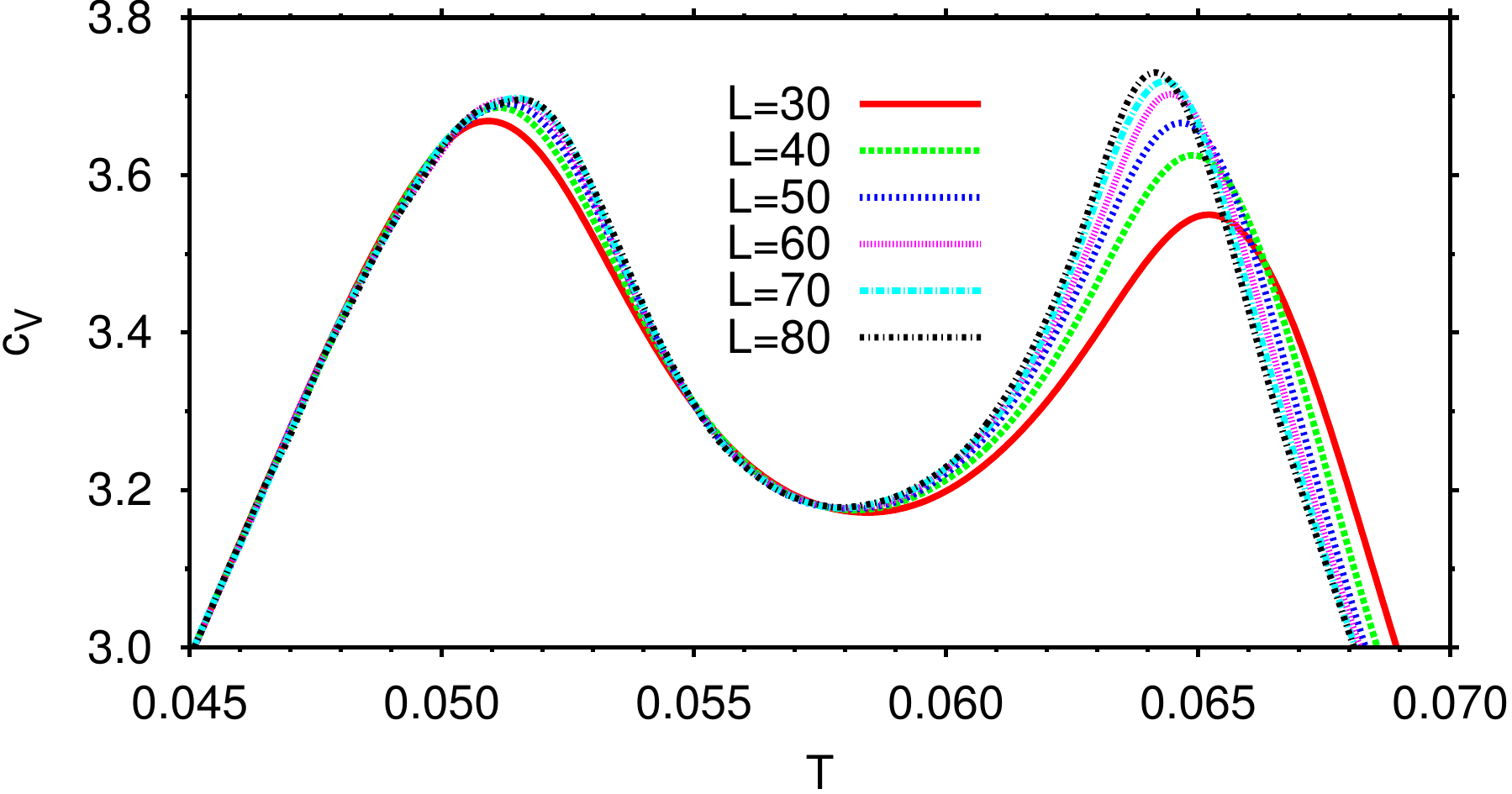}
\caption{\label{fig1} Specific heat $c_V$ of the BVST model, versus the 
temperature $T$, for various system sizes $L$. The data reveal 
two maxima, consistent with the two-transition scenario of the $q=6$ clock 
model. Note that finite-size effects in the peak heights are weak, indicating 
that $c_V$ does not diverge at any of the transitions.}
\end{center}
\end{figure}

In \fig{fig1}, we plot the specific heat per particle of the BVST model, $c_V = 
(\avg{E^2} - \avg{E}^2)/(NT^2)$, versus the temperature $T$, for various system 
sizes $L$. Consistent with the $q=6$ clock model, the data strikingly reveal two 
maxima, corresponding to two phase transitions. In addition, finite-size effects 
in $c_V$ are small. The absence of a strong increase of the peak heights with 
$L$ indicates that $c_V$ does not diverge at any of the transitions, consistent 
with the $q=6$ clock model. The reader is encouraged to compare our \fig{fig1} 
to specific heat MC data of the $q=6$ clock 
model~\cite{tobochnikPropertiesQStateClock1982, 
challaCriticalBehaviorSixstate1986}, which look very similar.

\subsection{Susceptibility}

\begin{figure}
\begin{center}
\includegraphics[width=\columnwidth]{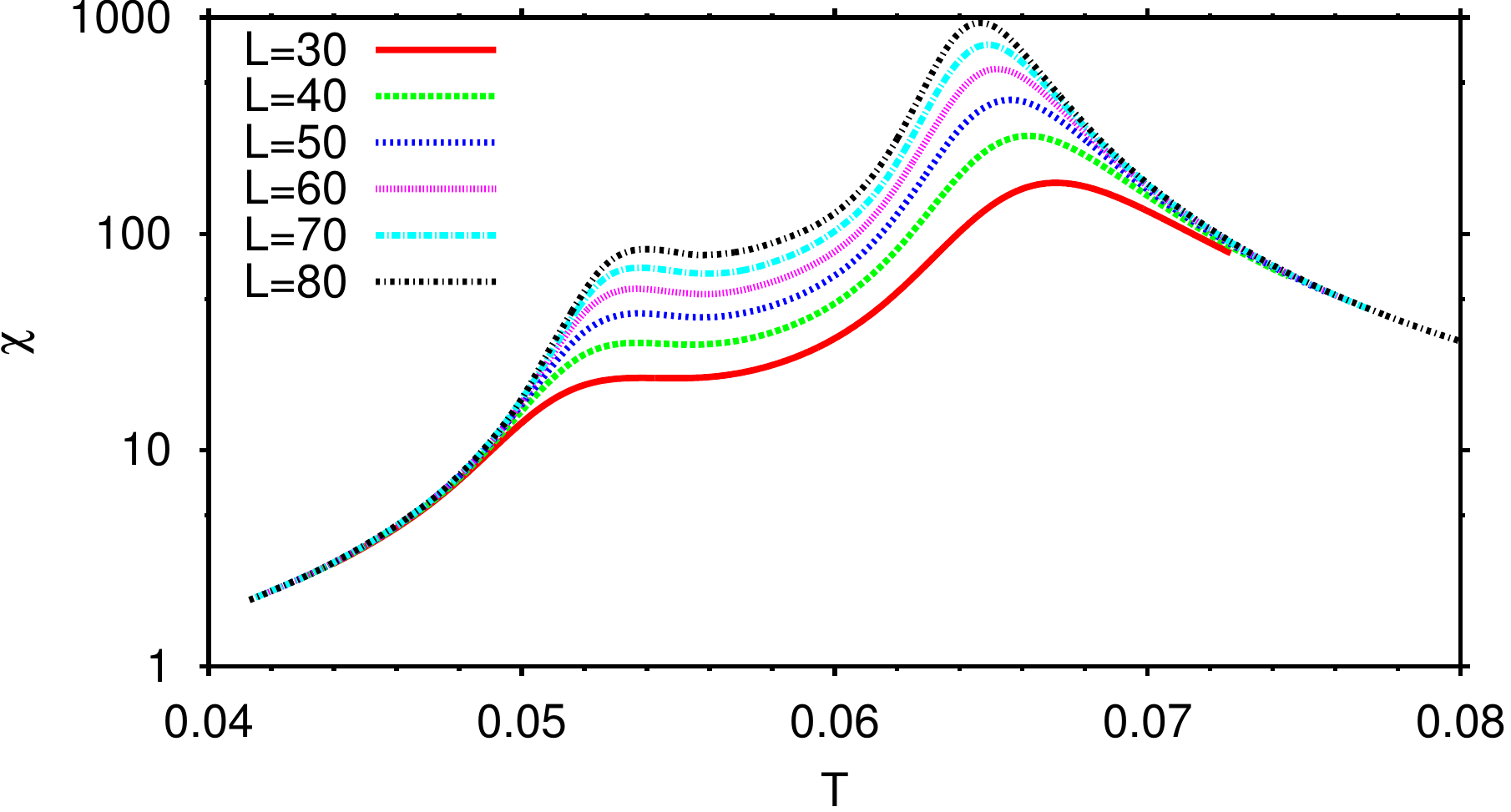}

\caption{\label{fig2} Susceptibility $\chi$ of the BVST model, versus the 
temperature $T$, for various system sizes $L$. The data reveal 
negligible size effects at low and high temperature, while $\chi$ increases 
strongly with $L$ at intermediate temperatures (note the vertical logarithmic 
scale). This supports the two-transition scenario of the $q=6$ clock model.}

\end{center}
\end{figure}

For a given set of particle displacements, in line with 
\olcite{citeulike:13839098}, we use the vector sum, $M = | \sum_{i=1}^N 
\vec{r}_i | / a_0$, to quantify the degree of order. In the perfectly ordered 
ground state $M/N=1$, since here all the particles have selected the same 
minimum, while $M/N<1$ when the ordering is not perfect. In the language of 
vector spin models, $M$ is analogous to the absolute value of the total 
magnetization, commonly used in studies of the $q=6$ clock 
model~\cite{tobochnikPropertiesQStateClock1982, 
challaCriticalBehaviorSixstate1986}. Note, however, that this is not the only 
possible choice, for example, an angular order parameter could be used 
also~\cite{baekTrueQuasilongrangeOrder2009}. We now define the susceptibility, 
$\chi = \left( \avg{M^2} - \avg{M}^2 \right)/ (NT)$, which we plot in \fig{fig2} 
as function of temperature, for various system sizes. The key observations are 
that finite-size effects are negligible in the low- and high-temperature 
regimes, while they are very strong in the intermediate regime, roughly 
corresponding to the temperature range spanned by the specific heat maxima. The 
pronounced increase of $\chi$ with $L$ at intermediate temperatures is 
consistent with a critical phase, where $\chi \to \infty$ in the thermodynamic 
limit $L \to \infty$. Hence, \fig{fig2} supports the two-transition scenario of 
the $q=6$ clock model, with non-critical phases at low and high temperature, 
separated by a critical phase (although we still need to check the nature of the 
order in each of the phases).

\begin{figure}
\begin{center}
\includegraphics[width=\columnwidth]{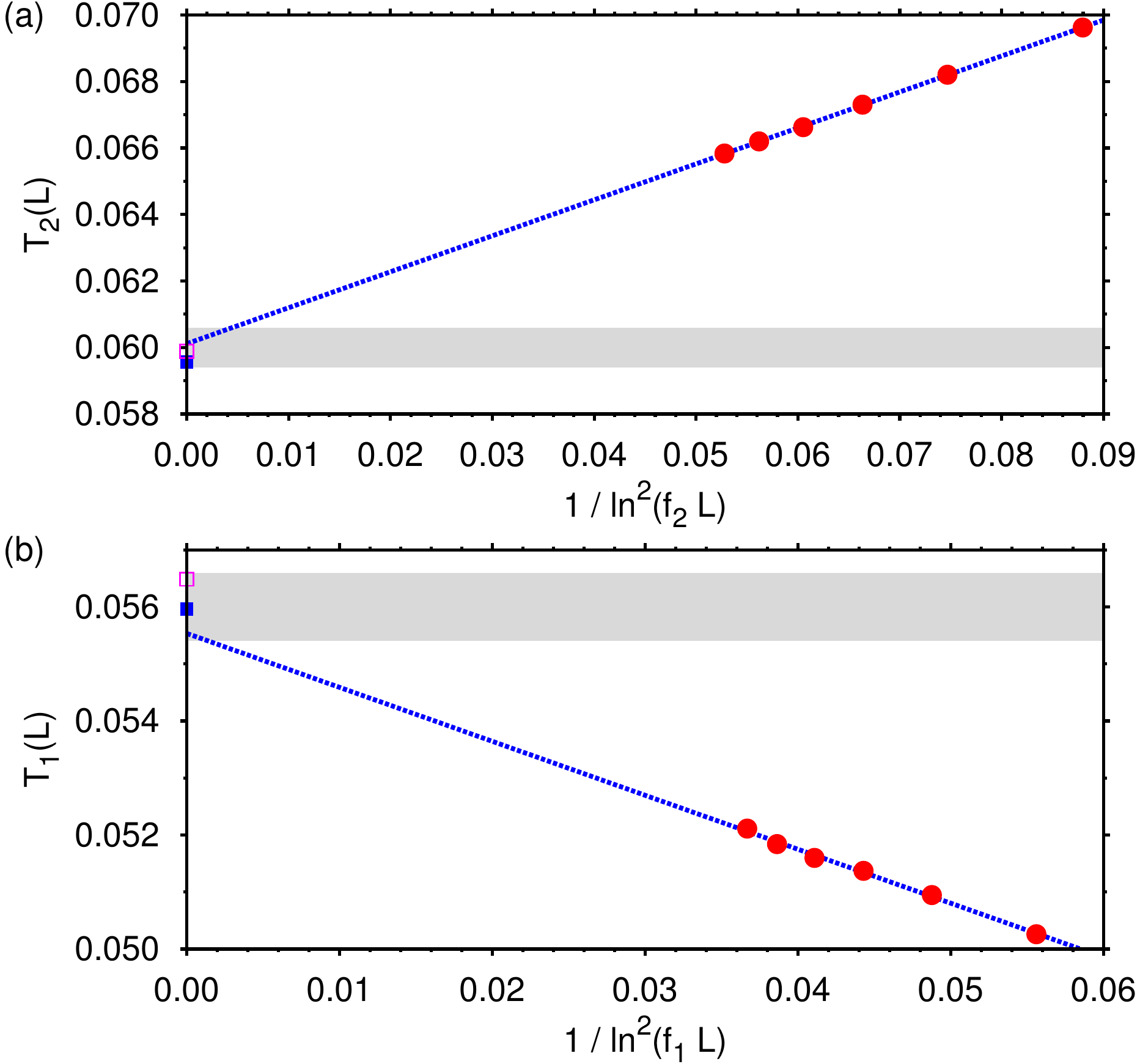}

\caption{\label{fig3} Finite-size scaling analysis of the temperatures (a) 
$T_2(L)$, and (b) $T_1(L)$, assuming the transitions are of the KT type. The 
intercept of the lines yields the transition temperature of the thermodynamic 
limit when all three parameters are fitted. The square symbols indicate the 
transition temperature obtained for two-parameter fits, where $a_1=a_2=a_{XY}$ 
was imposed, using all system sizes (closed symbol), and system sizes $L=70,80$ 
(open symbol).}

\end{center}
\end{figure}

\subsection{Phase transition temperatures}

To determine the transition temperatures, we perform a finite size scaling 
analysis. In a system of size $L$, approaching the lower transition at $T_1$ 
from below, \fig{fig2} shows that the slope $d\chi/dT$ initially increases, then 
levels off, i.e.~$d\chi/dT$ reaches a local maximum. Let $T_1(L)$ be the 
temperature at the maximum. Similarly, approaching the upper transition at $T_2$ 
from above, $d\chi/dT$ reaches a local minimum, defining $T_2(L)$. In the limit 
$L \to \infty$, these finite-size estimators converge to the transition 
temperature of the thermodynamic limit: $\lim_{L \to \infty} T_i(L)=T_i$ 
($i=1,2$).

Assuming the $q=6$ clock model scenario, the transitions at $T_1,T_2$ are both 
of the KT type, with $\xi$ given by \eq{eq:kt}. This implies scaling relations 
$T_1(L) = T_1 - a_1^2 / \ln^2(f_1 L)$ and $T_2(L) = T_2 + a_2^2 / \ln^2(f_2 L)$, 
with $f_i$ constants of order unity, and $a_i$ the coefficients of 
\eq{eq:kt}~\cite{chungEssentialFinitesizeEffect1999}. In \fig{fig3}(a), we fit 
our $T_2(L)$ estimates to the expected scaling form. The fit captures the data 
well, using $T_2 \approx 0.060$, $a_2 \approx 1.3$, $f_2 \approx 1.0$. 
\fig{fig3}(b) shows the corresponding fit to our $T_1(L)$ data, where $T_1 
\approx 0.056$, $a_1 \approx 1.3$, $f_1 \approx 2.3$ yielded the best fit. It is 
gratifying that both fits yield the same value for $a_i$, although the expected 
KT value is somewhat higher. If we repeat the procedure setting $a_1=a_2=a_{XY} 
\sim 1.5$, and only fit $f_i,T_i$, we obtain good fits also. The closed square 
symbols in \fig{fig3} indicate the transition temperatures obtained in this way, 
with $f_1 \approx 3.7$, $f_2 \approx 1.3$. Finally, again setting 
$a_1=a_2=a_{XY} \sim 1.5$, but this time only using data for $L=70,80$, i.e.~our 
largest systems, the open square symbols are obtained. From this analysis, we 
conclude $T_1 = 0.0560 \pm 0.0006$, and $T_2=0.0600 \pm 0.0006$; the vertical 
height of the shaded bands in \fig{fig3} indicates the corresponding ranges.

\subsection{Critical exponents}

\begin{figure}
\begin{center}
\includegraphics[width=\columnwidth]{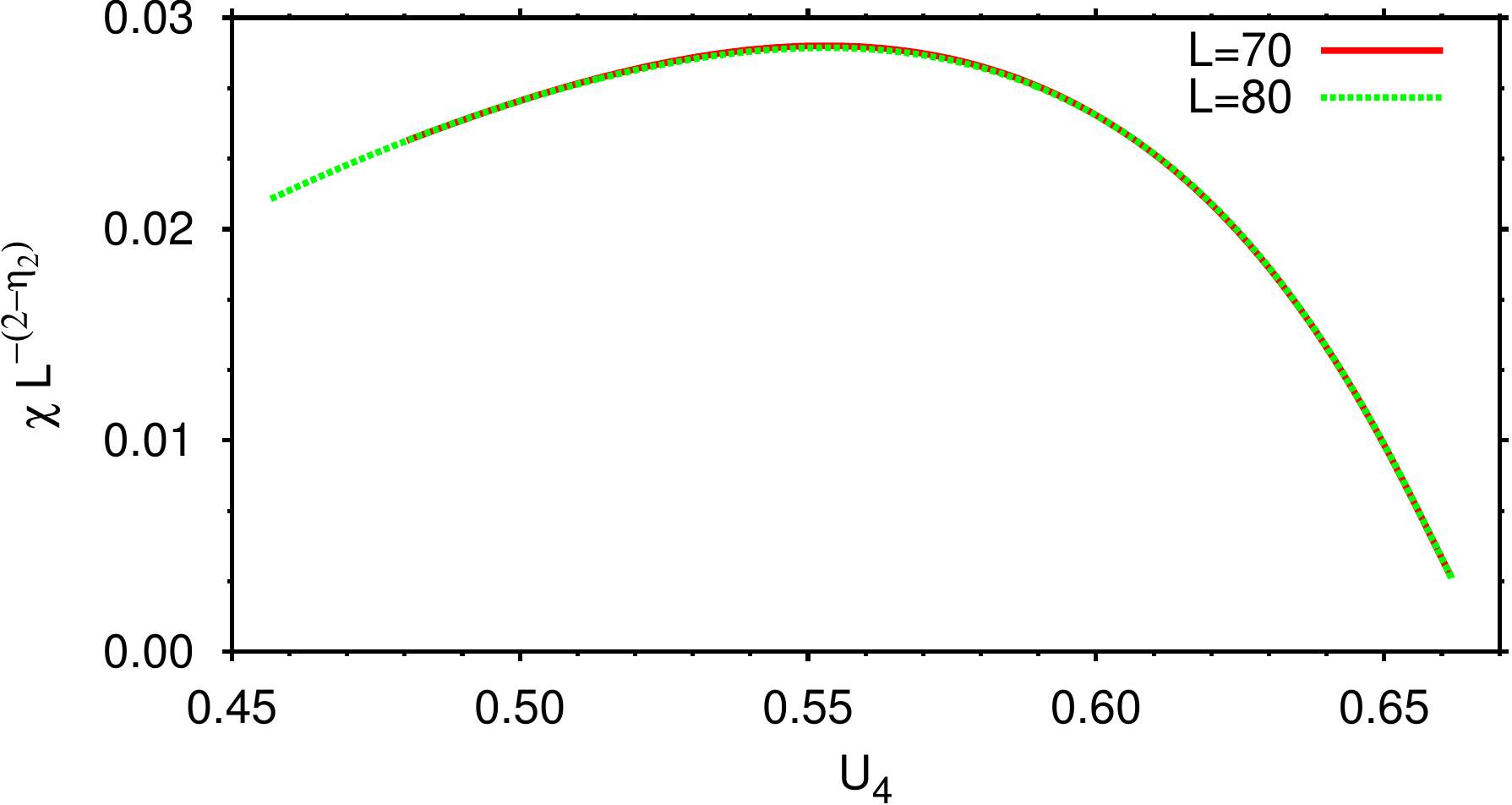}

\caption{\label{fig4} Application of the method of 
Loison~\cite{loisonBinderCumulantKosterlitzThouless1999} to obtain the critical 
exponent $\eta_2$ of the high-temperature transition. For this analysis, only 
data in the high-temperature phase, $0 < t_2 < 0.1$, was used, with $t_2$ 
defined in \eq{eq:kt}. Substituting the accepted value $\eta_2=1/4$ an excellent 
collapse of the data is observed!}

\end{center}
\end{figure}

Next, we turn to measuring the critical exponent $\eta_2$, i.e.~the value of 
$\eta$ at the high-temperature transition. The latter is most conveniently 
obtained using the method of 
Loison~\cite{loisonBinderCumulantKosterlitzThouless1999}; see also 
\olcite{baekTrueQuasilongrangeOrder2009} where this method is applied to the 
$q=8$ clock model. Here, one varies $T$, and plots the scaled susceptibility, 
$\chi L^{-(d-\eta_2)}$, with $d=2$ the spatial dimension, as function of the 
Binder cumulant, $U_4 = 1 - \avg{M^4} / (3\avg{M^2}^2)$. Provided the correct 
value of $\eta_2$ is used, the data for different system sizes $L$ should 
collapse onto a single curve. The result is shown in \fig{fig4}, where the 
accepted value $\eta_2=1/4$ was substituted, using our largest set of system 
sizes $L=70,80$. The quality of the collapse is quite remarkable! If we repeat 
the analysis using all our available system sizes, a somewhat larger exponent, 
$\eta_2 \sim 0.3$, is obtained. In \olcite{baekTrueQuasilongrangeOrder2009}, it 
is mentioned that deviations are likely due to logarithmic size corrections, 
which still are strong in small system sizes. Hence, we believe that 
$\eta_2=1/4$ obtained for our largest systems is the most reliable estimate, 
fully consistent with the $q=6$ clock model.

\begin{figure}
\begin{center}
\includegraphics[width=\columnwidth]{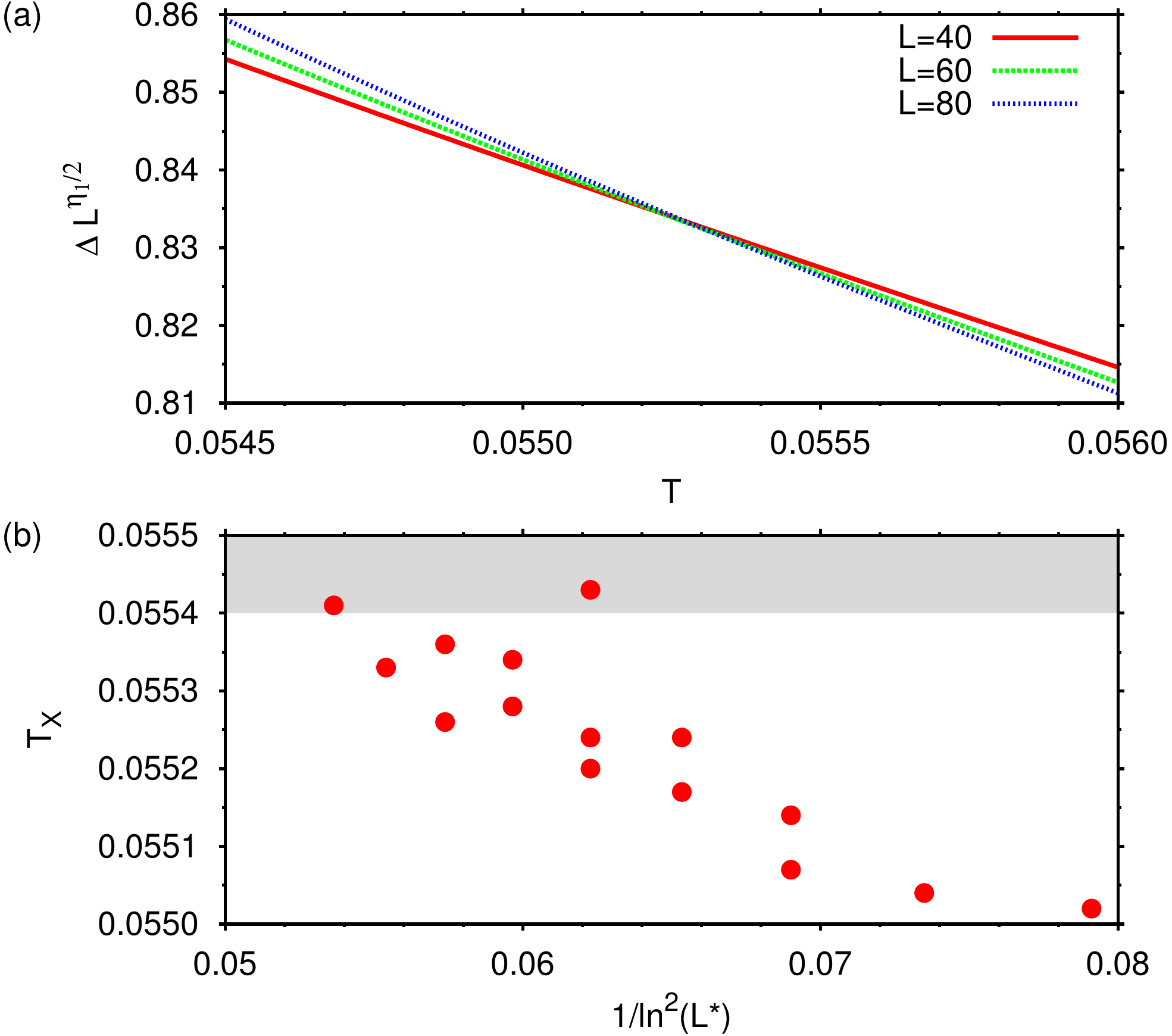}

\caption{\label{fig5} Finite-size scaling analysis of the order parameter 
$\Delta$ in the vicinity of the low-temperature transition. (a) $\Delta 
L^{\eta_1/2}$ versus $T$ for three different system sizes $L$, where 
$\eta_1=1/9$ was used. Note that curves for different $L$ intersect. (b) 
Intersection temperatures $T_X$ for pairs of system sizes, plotted versus 
$1/\ln^2 L^\star$. The shaded region marks the lower portion of the $T_1$ 
estimate range of \fig{fig3}(b). As $L^\star$ increases, $T_X$ approaches this 
range.}

\end{center}
\end{figure}

For the low-temperature transition, Loison's method is not directly applicable, 
and so, to obtain $\eta_1$, we follow a different route, based on the order 
parameter, $\Delta = \avg{M}/N$. In the critical intermediate phase, i.e.~for 
temperatures $T_1 \leq T \leq T_2$, the order parameter vanishes, following 
finite-size scaling, as $\Delta \propto L^{-\eta(T)/2}$, where the exponent 
$\eta(T)$ is an increasing function of $T$. For $T<T_1$, i.e.~in the ordered 
phase, $\Delta$ is finite, since here there is long-ranged order. Consider now 
the quantity $\Delta L^{\eta_1/2}$, with $\eta_1=1/9$ being the expected 
theoretical value of $\eta(T)$ at the lower transition temperature $T_1$. This 
quantity should diverge with $L$ for $T<T_1$ (since here $\Delta$ is finite), 
remain constant at $T_1$, and decay to zero above $T_1$ (since $\eta(T)$ 
increases with $T$). Hence, plotting $\Delta L^{\eta_1/2}$ versus $T$, for 
different system sizes $L$, the data for different $L$ are expected to intersect 
at $T=T_1$. The result is shown in \fig{fig5}(a), which clearly reveals 
intersections! For a pair of system sizes, $(L_i,L_j)$, we now define $T_X$ as 
the temperature where the corresponding curves intersect. \fig{fig5}(b) shows 
$T_X$ versus $1/\ln^2(L^\star)$, $L^\star \equiv (L_i+L_j)/2$, for all available 
pairs ($L=30,40,50,60,70,80$, i.e.~a total of 15 pairs). As $L^\star$ increases, 
there is a clear trend for $T_X$ to increase as well, approaching values that 
are consistent with \fig{fig3}(b). Hence, while \fig{fig5} does not constitute a 
direct measurement of $\eta_1$, it does show that the BVST model is {\it 
consistent} with the theoretically expected $q=6$ clock model value 
$\eta_1=1/9$.

\subsection{Correlation functions}

\begin{figure}
\begin{center}
\includegraphics[width=\columnwidth]{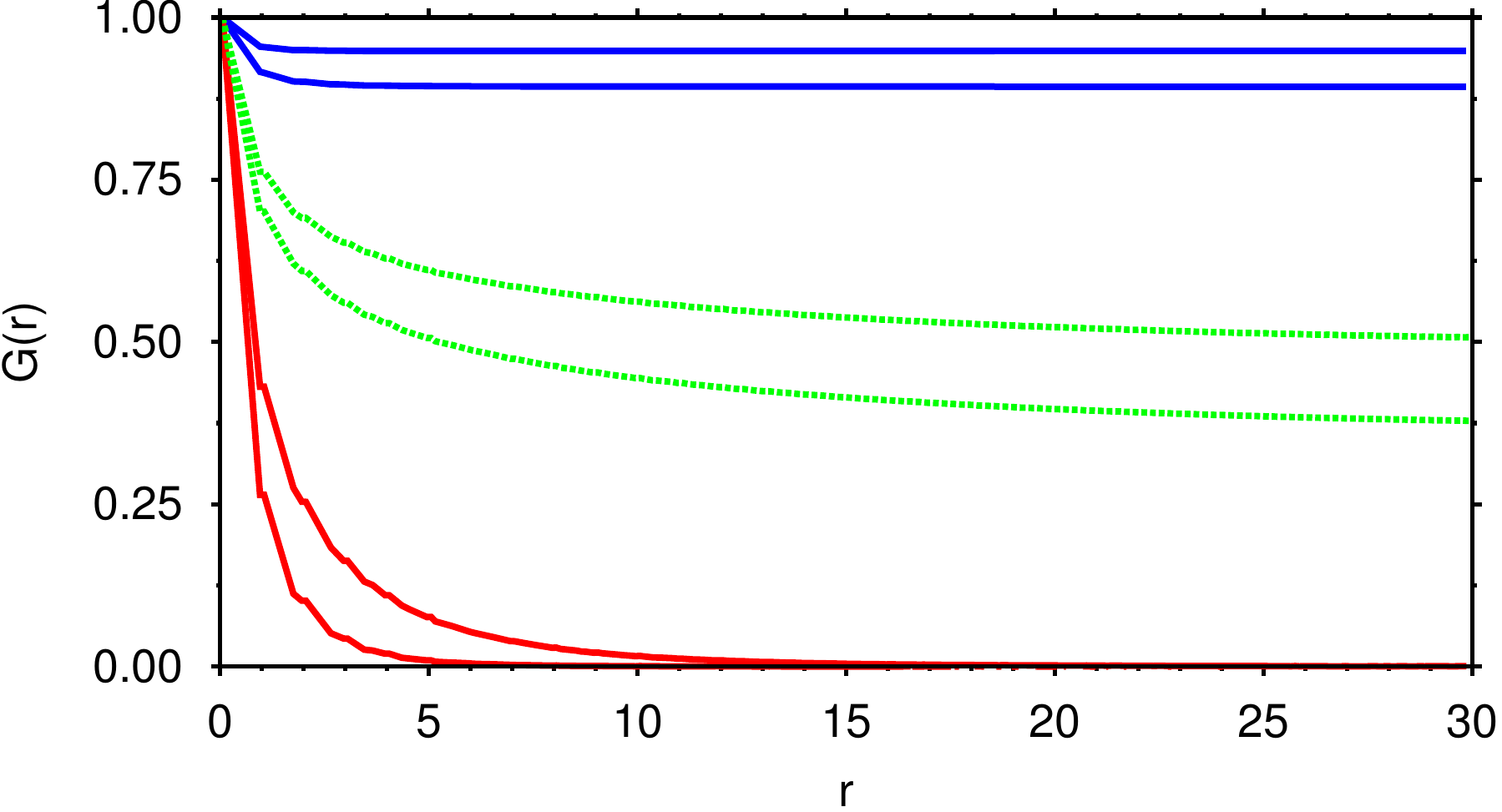}

\caption{\label{fig6} Correlation functions, $G(r)$, of the BVST model 
obtained at $T=0.04; 0.045$ (top two curves), $T=0.056; 0.06$ (center two 
curves), $T=0.075; 0.1$ (lower two curves). The three sets of curves represent, 
from top to bottom, the low-temperature phase with long-ranged order, the 
critical intermediate phase with power-law decay of correlations, and the 
high-temperature phase with short-range order.}

\end{center}
\end{figure}

Finally, we still present the correlation function $G(r)$ [\fig{fig6}]. For 
$T<T_1$, one indeed finds that $G(r)$ decays to a finite plateau value, 
consistent with long-ranged order, and a finite correlation length $\xi$. For 
$T>T_2$, $G(r)$ quickly decays to zero, consistent with short-ranged order, and 
finite $\xi$. In the intermediate phase, $G(r)$ decays slowly, but there is no 
sign of $G(r)$ saturating to a finite value, consistent with a critical phase 
where $\xi$ is infinite. Hence, also the dependence of $G(r)$ with temperature 
is consistent with the $q=6$ clock model.

\section{Conclusion}

\begin{figure}
\begin{center}
\includegraphics[width=\columnwidth]{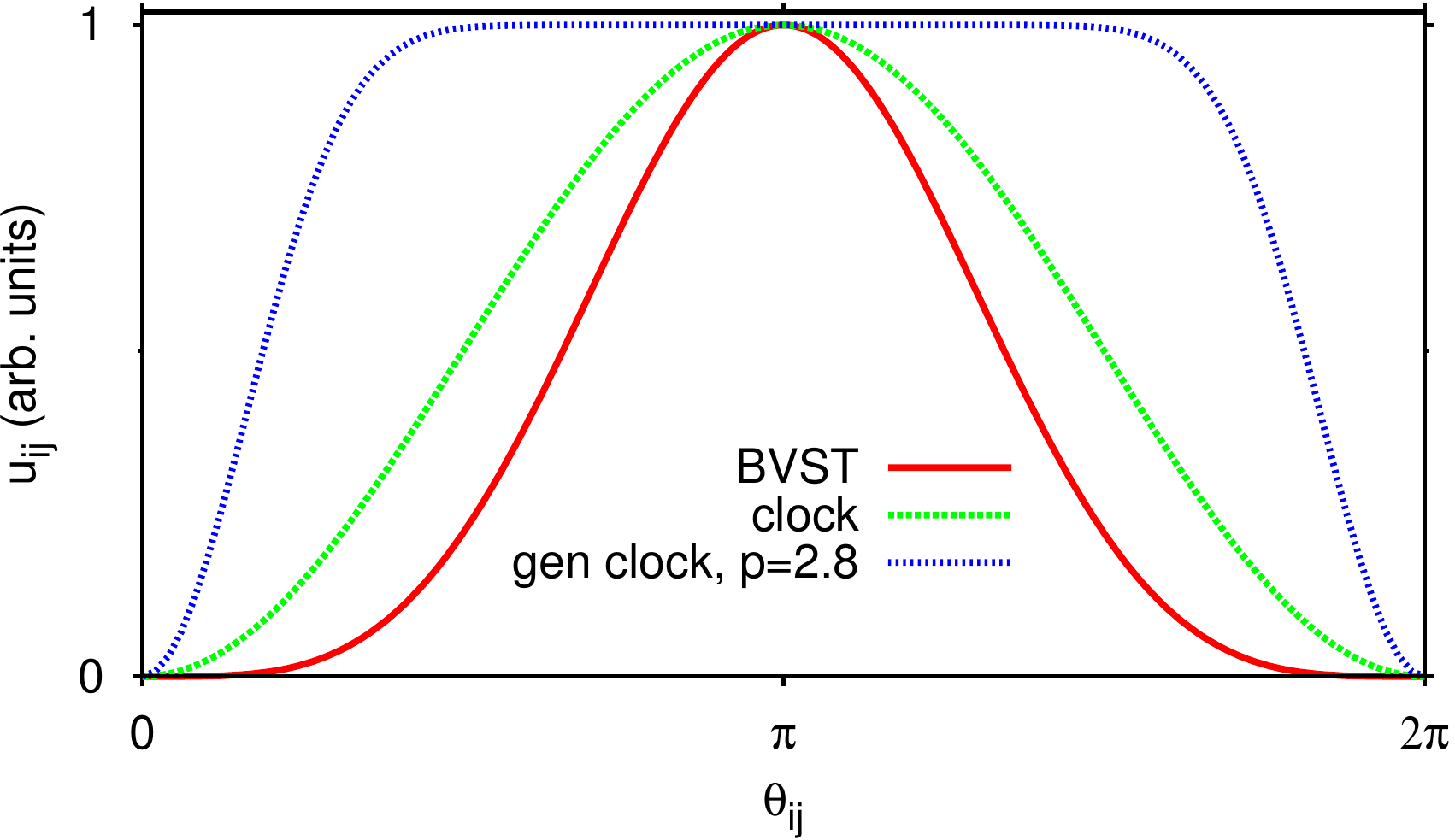}

\caption{\label{fig7} Shape of the nearest-neighbor pair interaction for the 
BVST model, the clock model, and the generalized clock model using $p=2.8$. In 
this plot, the potential minimum of each interaction was shifted to zero, 
followed by a scaling of the potential to have the maximum at unity.}

\end{center}
\end{figure}

In conclusion, we have presented MC data of the BVST model, showing that it 
belongs to the universality class of the two-dimensional $q=6$ clock model. The 
hallmark features are the presence of two KT transitions with, consequently, 
three thermodynamically distinct phases. Our data are consistent with the 
theoretical $q=6$ clock model critical exponents $\eta_1=1/9$ and $\eta_2=1/4$, 
of the lower and higher transition, respectively. Our estimates of the 
transition temperatures, $T_1$ and $T_2$, are somewhat below $\tc \approx 0.075$ 
reported in \olcite{citeulike:13839098}. However, since the focus of 
\olcite{citeulike:13839098} was on frictional behavior, no finite-size scaling 
study was performed, which we believe explains the deviation.

The reason the BVST model is in the $q=6$ clock model universality class is due 
to the external field $h_i$ {\it and} the bond potential $u_{ij}$. For each 
particle, the field gives rise to six local energy minima for this particle, 
corresponding to the \ahum{clock} states. It is important to note that, already 
at temperatures $T \sim T_2$, the external field is \ahum{confining} each 
particle to its set of minima. Hence, even though each particle in the BVST 
model has two degrees of freedom, $\vec{r}_i=(x_i,y_i)$, the field effectively 
reduces this to a set of six possible angles, $\theta_i=\pi\lambda/3$, 
$|\vec{r}_i|=a_0$, $\lambda=1,\ldots,6$, just as in the clock model. If we now 
plot the BVST bond energy, $u_{ij}$, as a function of the angular difference 
$\theta_{ij}$ between two nearest-neighboring particles, the bell-shaped curve 
\ahum{BVST} of \fig{fig7} is obtained, which qualitatively resembles the 
\ahum{cosine} shape of the pair potential of the clock model. Hence, in the BVST 
model, both the number of states and the shape of the pair potential correspond 
to the $q=6$ clock model, which explains why the universality classes are the 
same.

We still comment on the possibility of a single transition. Our data indicate 
that the difference in transition temperatures, $T_2-T_1$, is rather small. 
Hence, a critical reader might argue that, in the limit $L \to \infty$, the two 
transitions could well merge into one. However, based on what is known about 
clock models and their generalizations, this scenario is unlikely. The 
above-mentioned merging of transitions is known to happen in {\it generalized} 
clock models, where the nearest-neighbor interaction $u_{ij} \sim 1 - 
\cos^{2p^2} (\theta_{ij}/2)$, which reduces to the standard clock model when 
$p=1$. The two transitions merge into a single transition when the 
nearest-neighbor interaction becomes sufficiently \ahum{sharp and narrow}, 
implying a large enough value of $p$. For the $q=8$ clock model, the 
corresponding value $p>2.8$~\cite{baekTrueQuasilongrangeOrder2009}. However, as 
one can see in \fig{fig7}, the BVST pair potential is far removed from the 
\ahum{sharp and narrow} shape required to bring about such merging (on the 
contrary, the BVST model rather resembles $p<1$). As a side remark, if the 
transitions were to merge, we should expect two-dimensional $q=6$ state Potts 
behavior~\cite{baekTrueQuasilongrangeOrder2009}. The latter has a first-order 
transition~\cite{revmodphys.54.235}, for which neither our data, nor that of 
\olcite{citeulike:13839098}, show any evidence.

Finally, we discuss what might be expected in $d=3$ dimensions. In $d=3$, the 
clock model has a single second-order phase transition, for all values of 
$q$~\cite{scholtenCriticalBehaviorQstate1993a}. For $q=6$, the critical 
exponents are consistent with those of the $d=3$ 
XY-model~\cite{miyashitaNatureOrderedPhase1997}. Hence, the two-transition 
scenario of $d=2$ does not survive in $d=3$. We emphasize that, in the $d=3$ 
clock/XY models, only the lattice space is three-dimensional, the angular 
degrees of freedom remain two-dimensional. A more realistic description of a 
displacive transition in $d=3$ should likely use three-dimensional displacement 
vectors. For an fcc-material, a possible generalization of the BVST model could 
be an fcc-lattice, where, to each lattice site, 12 minima are assigned, each one 
displaced a small distance in the direction of one of the nearest neighbors. 
Such a model could easily be simulated using Monte Carlo methods, but we are not 
aware of such simulations having been carried out.

\acknowledgments

We acknowledge support by the German research foundation (SFB-1073, TP A01). We 
are additionally indebted to an anonymous referee for bringing to our attention 
the two-transition scenario of the six-state clock model, and for pointing out 
its possible connection to the BVST model.

\bibliographystyle{clear}
\bibliography{zot_home}

\appendix

\section{Monte Carlo methods}

\begin{figure}
\begin{center}
\includegraphics[width=\columnwidth]{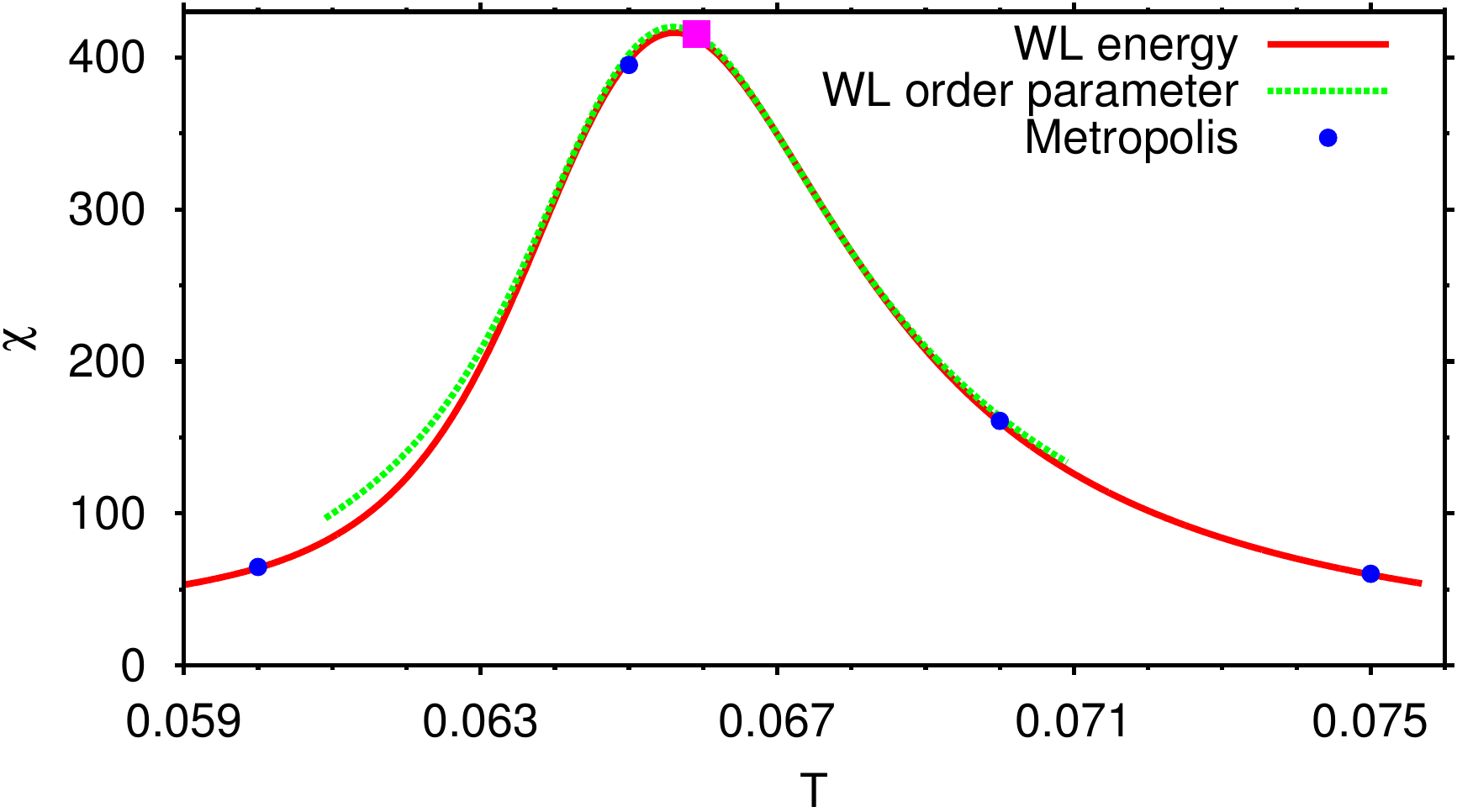}

\caption{\label{fig8} Susceptibility $\chi$ versus temperature $T$, for system 
size $L=50$, obtained using three different MC methods. There is overall good 
agreement between WL energy sampling and the Metropolis method. For WL order 
parameter sampling, the simulation temperature that was used is marked with the 
square symbol. In the vicinity of this temperature, there is good agreement with 
the other methods, but agreement deteriorates as one moves away from this 
temperature.}

\end{center}
\end{figure}

We simulate the BVST model using various MC techniques. The principal MC move is 
always the random selection of a single particle, which then gets translated by 
a (two-dimensional) vector drawn randomly from a circle of radius $r/a=0.03$. 
The use of a single-particle move restricts us to moderate system sizes, owing 
to critical slowing down, but these nevertheless suffice to demonstrate the 
connection to the $q=6$ clock model, our main conclusion. To overcome critical 
slowing down, a cluster move for the BVST model is required, the development of 
which is beyond the scope of this work. For the determination of the correlation 
functions [\fig{fig6}] standard Metropolis sampling was used, where each MC move 
is accepted with probability $P = \min \left[1, \exp(-\Delta E / k_BT) \right]$, 
with $\Delta E$ the energy difference between initial and proposed state, and 
$T$ the temperature. For the determination of thermodynamic quantities of 
interest $(c_V,\chi,\Delta, U_4,\ldots)$ as function of $T$, we used Wang-Landau 
(WL) {\it energy} sampling~\cite{wang.landau:2001, citeulike:278331}. In this 
method, the simulation performs a random walk on the energy range of interest, 
$0.05<E/N<0.17$, chosen such that both transitions are captured (the range was 
discretized in steps $\Delta E=0.25$). The principal output of these simulations 
is the density of states $g(E)$. Thermal averages may then be computed for any 
temperature of interest using $\avg{X} = Z^{-1} \sum_E X_E \, g(E) e^{-E/K_BT}$, 
where $X_E$ denotes the microcanonical average of~$X$, and normalization $Z = 
\sum_E g(E) e^{-E/K_BT}$. For example, to compute energy moments, $\avg{E^k}$, 
one sets $X_E=E^k$. To compute order parameter moments, $\avg{M^k}$, one sets 
$X_E = M_E^k$, defined as the average value of $M^k$ in the bin corresponding to 
energy~$E$. The latter quantities are readily collected during WL sampling by 
updating a small number of array elements after each MC move. In locating the 
temperatures $T_i(L)$ of \fig{fig3}, the slope $d\chi/dT$ needs to be computed. 
For better accuracy, we expressed $d\chi/dT$ in terms of appropriate moments of 
$E$ and $M$, as opposed to using a finite-difference scheme to differentiate 
$\chi$ directly (for example: $d\avg{M}/dT \propto \avg{ME}-\avg{M}\avg{E}$, and 
so forth). To further enhance efficiency, we complemented our WL simulations by 
the collection of transition matrix elements~\cite{citeulike:12476499, 
citeulike:2299400}, following the implementation of \olcite{citeulike:3577799}. 
In our simulations, the energy range of interest is split into $\sim 15-30$ 
intervals, with a single processor assigned to each interval (runtime per 
processor $\sim 44-88$ hours), the data being combined afterward. The accept 
rate of MC moves varies $\sim 10-40 \%$ (highest on the high-energy side), 
performance being $\sim 4 \times 10^6$ attempted moves per second. In the 
previous version of this work~\cite{vinkCriticalBehaviorDisplacive2018a}, WL 
sampling was performed over a specified {\it order parameter} range, see also 
\olcite{citeulike:13106236}. This method is advantageous for systems where 
reaching the ordered state is difficult, but has the disadvantage that the range 
in temperature, over which one can reliably compute thermal averages, is 
restricted. Since now two transitions need to be sampled, sampling over a 
specified energy range turned out to be the optimal choice. In \fig{fig8}, we 
present a comparison between measurements of the susceptibility obtained using 
all three methods.

\end{document}